\documentclass[journal]{IEEEtran}
\usepackage{epsfig,rotating,setspace,latexsym,amsmath,epsf,amssymb,amsfonts,bm,theorem,cite,authblk, bbm,color, hyperref, multirow, caption, graphicx, subcaption, cite, url}

\IEEEoverridecommandlockouts
\allowdisplaybreaks

\title{Privacy-Preserving Semantic Communications via Multi-Task Learning and Adversarial Perturbations}

\author[1]{Yalin E. Sagduyu}
\author[1]{Tugba Erpek}
\author[2]{Aylin Yener}
\author[3]{Sennur Ulukus}

\affil[1]{\normalsize  Nexcepta, Gaithersburg, MD, USA}

\affil[2]{\normalsize  The Ohio State University, Columbus, OH, USA}

\affil[3]{\normalsize University of Maryland, College Park, MD, USA}

\begin{document}
\maketitle
\thispagestyle{empty}

\begin{abstract}
Semantic communications conveys task-relevant meaning rather than focusing solely on message reconstruction, improving bandwidth efficiency and robustness for next-generation wireless systems. However, learned semantic representations can still leak sensitive information to unintended receivers (eavesdroppers). This paper presents a deep learning-based semantic communication framework that jointly supports multiple receiver tasks while explicitly limiting semantic leakage to an eavesdropper. The legitimate link employs a learned encoder at the transmitter, while the receiver trains decoders for semantic inference and data reconstruction. The security problem is formulated via an iterative min-max optimization in which an eavesdropper is trained to improve its semantic inference, while the legitimate transmitter-receiver pair is trained to preserve task performance while reducing the eavesdropper's success. We also introduce an auxiliary layer that superimposes a cooperative, adversarially crafted perturbation on the transmitted waveform to degrade semantic leakage to an eavesdropper. Performance is evaluated over Rayleigh fading channels with additive white Gaussian noise using MNIST and CIFAR-10 datasets. Semantic accuracy and reconstruction quality improve with increasing latent dimension, while the min-max mechanism reduces the eavesdropper's inference performance significantly without degrading the legitimate receiver. The perturbation layer is successful in reducing semantic leakage even when the legitimate link is trained only for its own task. This comprehensive framework motivates semantic communication designs with tunable, end-to-end privacy against adaptive adversaries in realistic wireless settings.
\end{abstract}

\section{Introduction}
Wireless systems increasingly support intelligent edge services, including classification, detection, and decision-making, under bandwidth and latency constraints. \emph{Semantic communications} enables these services by transmitting compact representations optimized for task success beyond reliable source reconstruction at the intended receiver \cite{gunduz2022beyond, uysal2021semantic}. Trained end-to-end over the wireless channel, semantic communication systems can learn latent representations robust to channel effects under limited channel uses, while enabling both semantic inference and reconstruction at the receiver \cite{Xie2021DeepLearningSemantic, sagduyu2024will}.

\emph{Privacy} has long been studied in information theory through fundamental limits on information leakage, commonly formalized via secrecy criteria and statistical measures \cite{bloch2021overview}. Complementary to these foundations, this paper focuses on \emph{semantic leakage} in deep learning-empowered semantic communications, where the risk goes beyond raw data recovery and focuses on task-relevant inference from learned representations.

\emph{Privacy concerns.} Representations optimized for meaning can still leak \emph{sensitive} information, since task-relevant features may enable semantic inference by an unintended eavesdropper \cite{guo2024survey}. Conventional physical-layer security primarily protects message bits from eavesdroppers, motivating privacy mechanisms that directly limit semantic leakage.

Several mechanisms have been studied for \emph{privacy-preserving semantic communications}. Eavesdropper recovery can be reduced using information-bottleneck representations with adversarial learning and adaptive weighting under channel variation \cite{wang2024privacy}. Knowledge-aware methods can mitigate leakage from shared or mismatched background knowledge via knowledge management, mapping, and disambiguation \cite{liu2025knowledge, cheng2024knowledge}. Confidentiality mechanisms include key-based encrypted semantic communication with adversarial encryption training, and comparative studies of perturbation, encryption, adversarially trained transceivers, and learning-based quantization for privacy-task-performance tradeoffs \cite{luo2023encrypted, guo2025semantic}. Domain-specific objectives (e.g., visual) can reduce the informativeness of intercepted reconstructions while preserving task performance \cite{Xu2023VisualProtection}. Deployment-oriented designs can combine semantic encryption, privacy-preserving perturbation, and destination-side semantic calibration \cite{liu2023semprotector}. 

\begin{figure}[h]
  \centering
    \includegraphics[width=\linewidth]{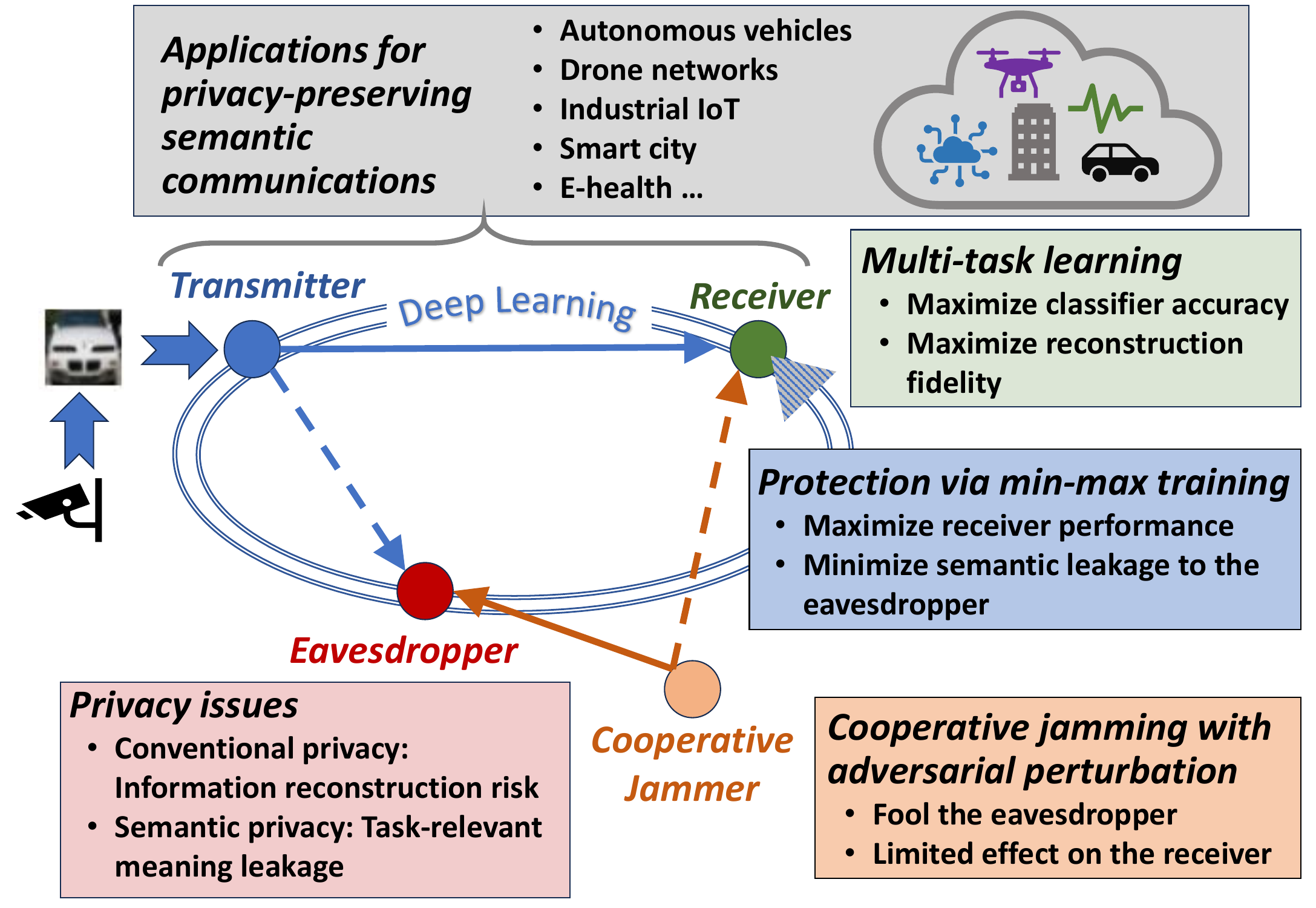}
    \caption{Privacy-preserving semantic communications: eavesdropping and protection mechanisms.}\label{fig:App}
\end{figure}

Building on these directions, this paper presents a \emph{unified end-to-end multi-task semantic communication framework} that enforces privacy through competitive min-max training against an eavesdropper equipped with adaptive learning, and introduces a complementary perturbation layer that reduces semantic leakage without modifying the core encoder-decoder objective. The eavesdropping threat and protection mechanisms for privacy-preserving semantic communications are illustrated in Fig.~\ref{fig:App}.

\emph{Semantic leakage protection via multi-task learning.} Intercepted learned representations motivate protection mechanisms that are native to semantic learning pipelines and tunable for privacy-utility tradeoffs against a strong adversary. Here, semantic leakage is scoped to the considered semantic task, and the eavesdropper's task accuracy is used as an empirical proxy for task-specific leakage. A \emph{min-max training} operation is applied in a \emph{multi-task learning} setting, shaping a single latent representation to jointly support semantic inference and reconstruction at the legitimate receiver while explicitly reducing an adaptive eavesdropper's success. The legitimate transmitter-receiver pair is trained for task accuracy, reconstruction quality, and protection, while the eavesdropper trains a semantic decoder for inference. This interaction, viewed as an \emph{adversarial min-max game} between the legitimate pair and the eavesdropper, is implemented via repeated \emph{best-response} updates. In each round, the eavesdropper strengthens its inference against the current transmitted representation, after which the legitimate pair preserves intended performance while reducing the eavesdropper's success. 

\emph{Semantic leakage protection via adversarial perturbations.} We also consider an auxiliary protection layer that adds a cooperative, \emph{adversarially crafted perturbation} to the transmitted waveform. While adversarial attacks have been studied in semantic communication systems \cite{sagduyu2022semanticsecurity}, here the perturbation is designed to selectively degrade the eavesdropper's inference while preserving legitimate task performance. The transmitter-receiver pair can be trained only for legitimate utility (without relying on access to eavesdropper's training process), and the perturbation is then designed to degrade the eavesdropper's inference with minimal impact on the legitimate link.

\emph{Performance-privacy tradeoffs and design guidelines.} Performance is evaluated on MNIST and CIFAR-10 datasets over Rayleigh-fading plus additive white Gaussian noise (AWGN) channels. The proposed multi-task semantic communication approach achieves high semantic accuracy and strong reconstruction quality measured by Peak Signal-to-Noise Ratio (PSNR) and Structural Similarity Index Measure (SSIM) with a limited number of channel uses. All performance metrics improve systematically as latent dimension or SNR increases.

Performance results show that augmenting legitimate training with min-max privacy objectives reduces semantic inference accuracy of an eavesdropper without degrading the legitimate receiver's semantic task performance. Stronger privacy weighting improves the privacy-utility tradeoff (including as channel uses increase), while the eavesdropper's performance improves with its SNR, yielding a tunable tradeoff against a continually strengthened eavesdropper. Privacy pressure is imposed while maintaining high semantic accuracy and only mildly affecting reconstruction at the legitimate receiver, and we characterize how latent dimension, as a bandwidth proxy, impacts this protection mechanism.

We further show that it is possible to reduce the eavesdropper's semantic inference capability even when the legitimate semantic link is trained without incorporating the eavesdropper. With the perturbation layer, single-step Fast Gradient Sign Method (FGSM) and iterative Projected Gradient Descent (PGD) methods are applied to generate adversarial perturbations that significantly reduce semantic leakage, with PGD providing stronger protection than FGSM and additional PGD steps yielding further improvement. 

Motivated by these design guidelines and observations, privacy can be treated as a \emph{tunable} objective in semantic communication systems and integrated directly into end-to-end learning without compromising task utility. This perspective encourages deployment-oriented architectures and training paradigms that preserve performance while providing robust, adaptive protection against increasingly capable eavesdroppers across heterogeneous operating conditions.

The remainder of the paper is organized as follows. Sec.~\ref{sec:sem} describes semantic communications with multi-task learning. Sec.~\ref{sec:privacy} introduces the eavesdropping threat model and presents the min-max training to reduce semantic leakage. Sec.~\ref{sec:perturbation} presents the perturbation-based protection mechanism against semantic leakage. Sec.~\ref{sec:future} outlines future research directions. Sec.~\ref{sec:conclusion} concludes the paper.

\section{Semantic Communications with Multi-Task Learning \label{sec:sem}}
\begin{figure}[ht]
  \centering
    \begin{subfigure}[t]{\columnwidth}
    \centering
    \includegraphics[width=\linewidth]{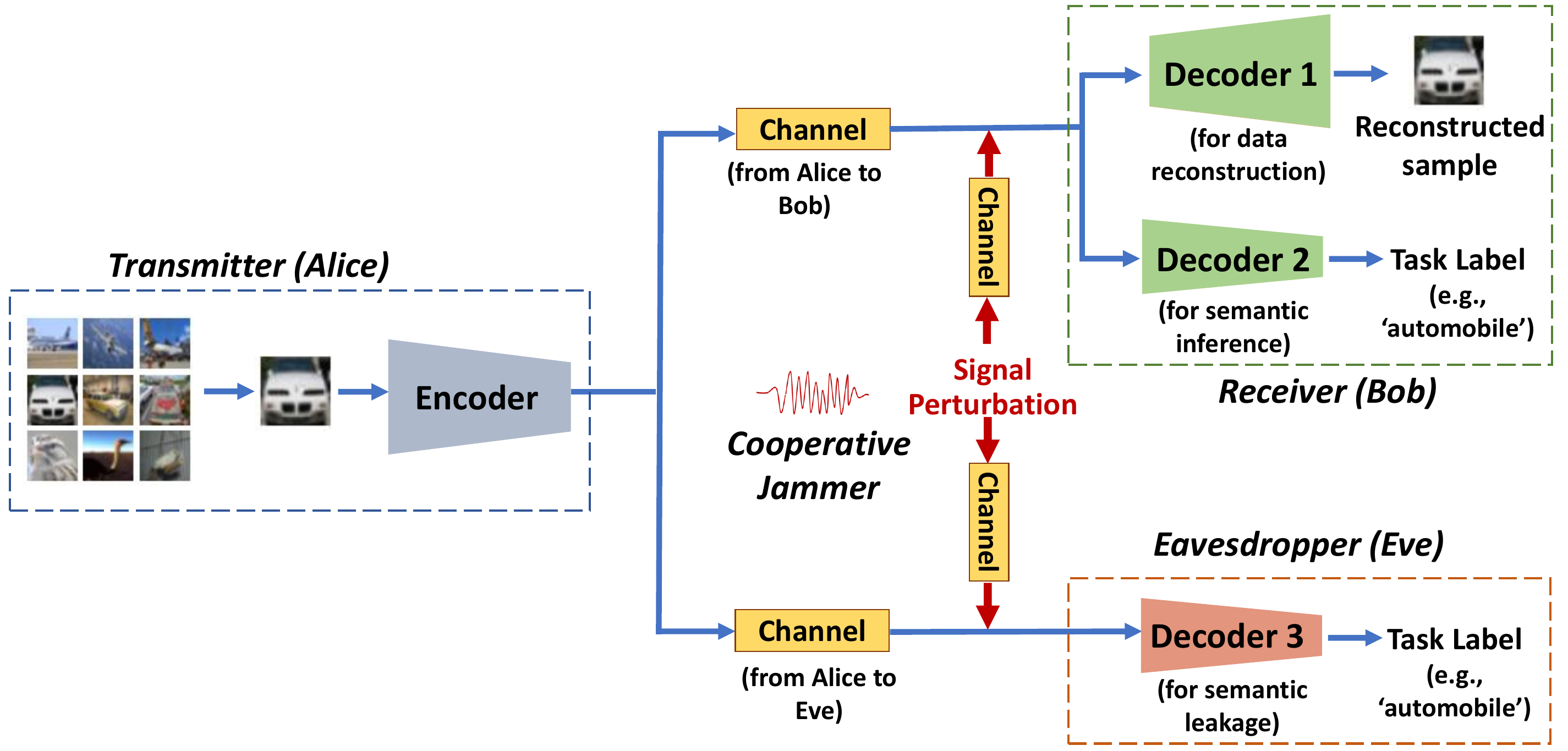}
    \caption{System model for privacy-preserving semantic communications.}
    \label{fig:SysMod}
  \end{subfigure}

    \vspace{0.6em}

  \begin{subfigure}[t]{\columnwidth}
    \centering
    \includegraphics[width=\linewidth]{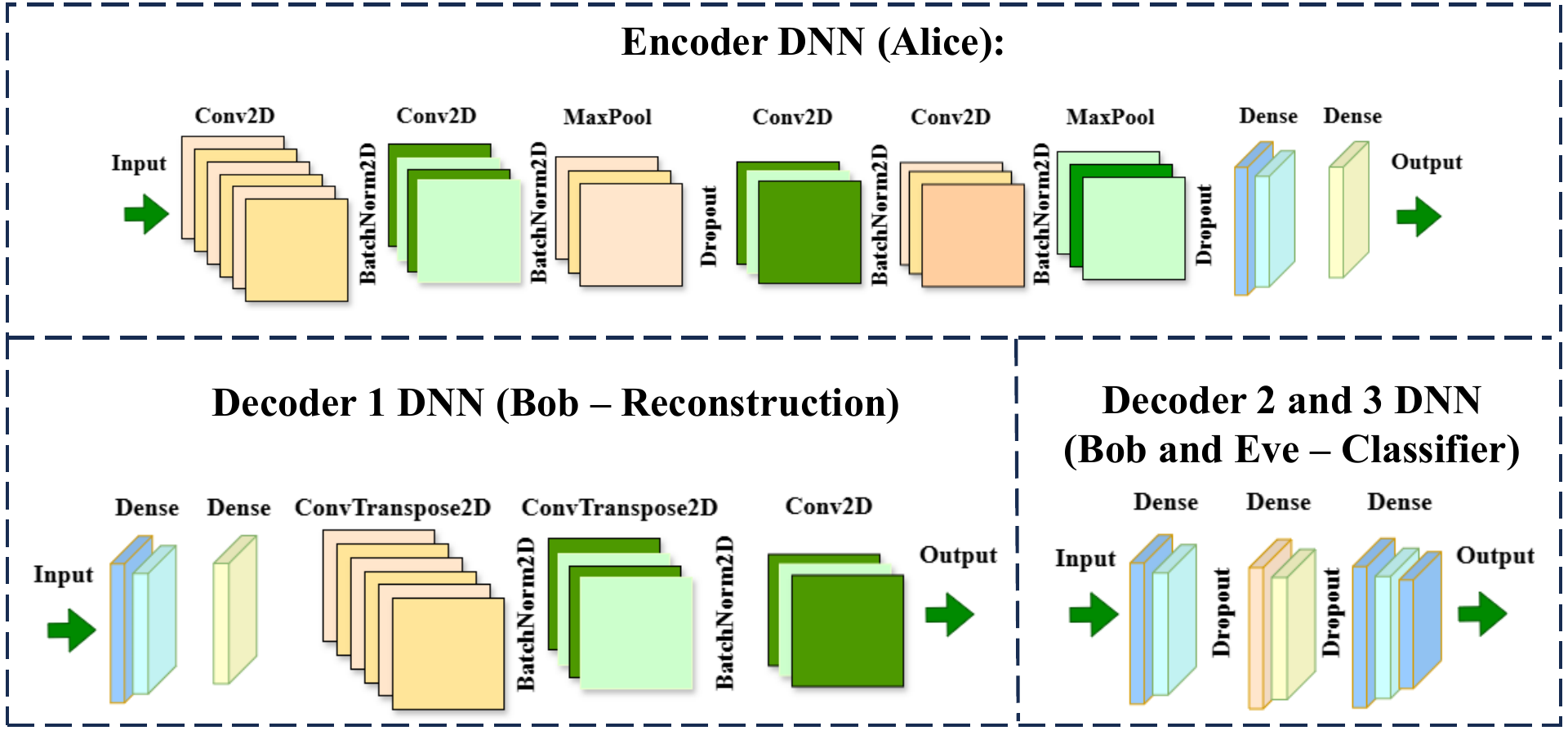}
    \caption{Neural network architectures for the encoder and decoders.}\label{fig:NNArch}
  \end{subfigure}

  \caption{System model for privacy-preserving semantic communications and underlying encoder and decoder architectures.}
  \label{fig:SysModNNArch}
\end{figure}

We consider an image source at the transmitter (Alice) and a legitimate receiver (Bob) operating over a wireless channel with Rayleigh fading and AWGN. The system model is shown in Fig.~\ref{fig:SysMod}. We assume that Alice and Bob each have a single antenna. The transmitter maps each input image into a fixed-length latent representation that corresponds to a limited number of complex channel uses represented by the dimension of the latent space. The wireless channel corrupts this representation through fading and noise, and the receiver performs learned decoding without requiring explicit channel state information (CSI).

\subsection{Encoder and Decoder at the Transmitter-Receiver Pair}
 Encoder and decoder architectures of Alice and Bob are shown in Fig.~\ref{fig:NNArch}. Alice uses a convolutional neural network (CNN)-based \emph{encoder} that compresses each input image into a compact latent representation for wireless transmission. The encoder is a two-stage convolutional feature extractor, where each stage uses two small kernel convolution layers with batch normalization and ReLU activations, followed by max-pooling-based downsampling and moderate dropout for regularization. The first stage expands the feature depth to 64 channels and the second expands it to 128 channels, reducing spatial resolution by an overall factor of four (e.g., to 7 by 7 for MNIST and 8 by 8 for CIFAR-10). The resulting feature map is flattened and passed through a two-layer fully connected projection head with a 1024-unit hidden layer to produce a concatenated in-phase and quadrature (I/Q) vector. This output is reshaped into two streams representing the in-phase and quadrature components over the chosen latent length (e.g., 32 or 256). Per-sample power normalization is applied at the encoder output so that each transmitted representation satisfies a consistent transmit power budget and the channel SNR interpretation remains stable across batches and epochs.

Bob uses a \emph{multi-task decoder} with two branches driven by the same received latent representation. \emph{The reconstruction branch} flattens the I/Q streams and applies a two-layer fully connected expansion (with a 1024-unit intermediate layer) to map the received vector into a low-resolution feature map with 256 channels at one-quarter spatial resolution. This feature map is then upsampled using two transposed-convolution blocks with batch normalization and ReLU, followed by a final convolution and sigmoid to produce valid pixel intensities. In parallel, \emph{the semantic branch} is a feed-forward classifier on the flattened I/Q vector with two hidden layers with ReLU and dropout, and a final linear layer that outputs class logits for the downstream task.

Training the two branches \emph{jointly} encourages the latent representation to preserve task-relevant meaning while also enabling monitoring of perceptual quality through reconstruction-based metrics. Bob is trained \emph{without explicit CSI} and robustness is learned end-to-end through exposure to the channel impairments during training. The SNR of the Alice-Bob channel is randomly varied over a broad range for each transmitted block during training, so that the encoder and Bob's decoders learn representations that generalize across the SNR conditions observed by Bob.

\subsection{Multi-task Learning Objective}
During training, Bob optimizes a combined objective for semantic correctness and reconstruction quality, implemented as a \emph{weighted sum of task losses}. The semantic loss is categorical cross entropy (CCE), while the reconstruction loss combines mean squared error (MSE) for distortion fidelity with an SSIM-based term for perceptual quality. At test time, performance is measured by classification accuracy, together with PSNR and SSIM for reconstruction quality. For images, PSNR and SSIM better capture perceptual fidelity and structural consistency than pixel-wise losses, with SSIM preserving task-relevant structure under semantic compression and PSNR providing a scale-normalized, interpretable MSE variant.

\emph{Multi-task learning} plays two complementary roles. First, it forces the transmitter representation to remain useful for inference, not only for pixel recovery. Second, it stabilizes learning under channel impairments: when semantic and reconstruction signals are both present, the encoder is less likely to overfit to a single objective that may be brittle under fading or noise. Practically, the semantic branch encodes meaning, while the reconstruction branch preserves structural information and improves interpretability. This multi-task baseline also provides a clear reference
point for the protection mechanism introduced later in Sec.~\ref{sec:privacy}, where an additional objective is imposed to suppress eavesdropper inference.

\subsection{Semantic Communications Performance}
Two image datasets are used to illustrate complexity effects. MNIST images are 28 by 28 pixels with a single grayscale channel, so their shape is 28$\times$28$\times$1. CIFAR-10 images are 32 by 32 pixels with three color channels, so their shape is 32$\times$32$\times$3. The reconstruction MSE loss is weighted by 5, while the other losses each use a weight of 1, to balance the relative contributions of reconstruction and semantic terms to the combined loss during joint training. Note that a tunable weight for the privacy protection loss is introduced in Section \ref{sec:privacy}. Training is performed in PyTorch for 500 epochs on NVIDIA RTX PRO 6000 Blackwell GPUs.

\begin{figure}[t]
  \centering
    \begin{subfigure}[t]{0.49\columnwidth}
    \centering
    \includegraphics[width=\linewidth]{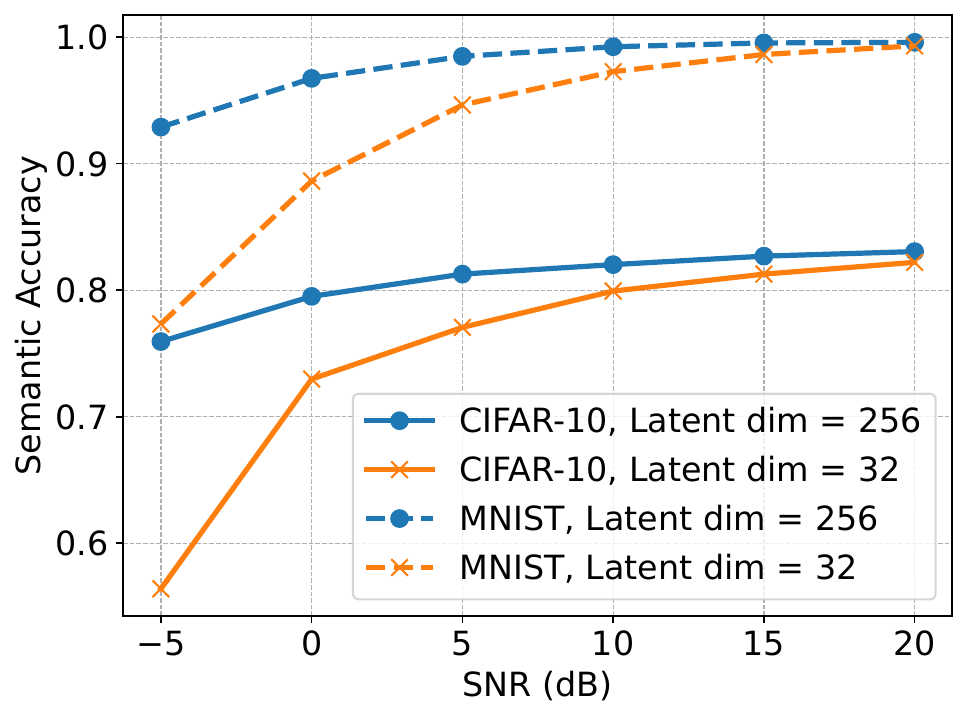}
    \caption{Semantic accuracy.}\label{fig:Sema}
  \end{subfigure}\hfill
  \begin{subfigure}[t]{0.49\columnwidth}
    \centering
    \includegraphics[width=\linewidth]{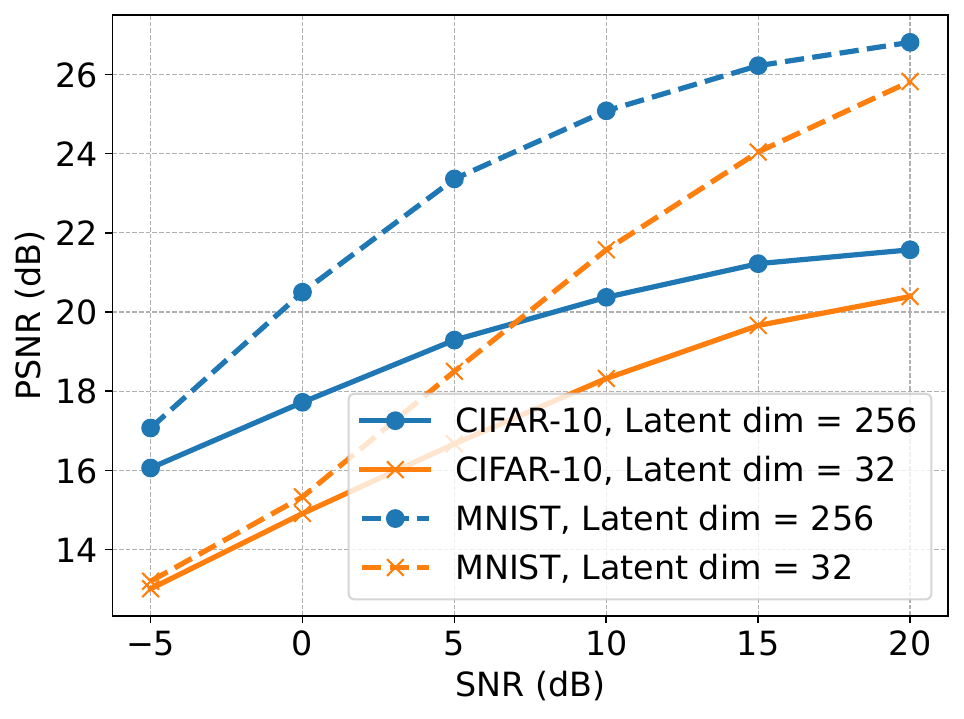}
    \caption{PSNR.}\label{fig:Semb}
  \end{subfigure}

    \vspace{0.6em}

  \begin{subfigure}[t]{0.49\columnwidth}
    \centering
    \includegraphics[width=\linewidth]{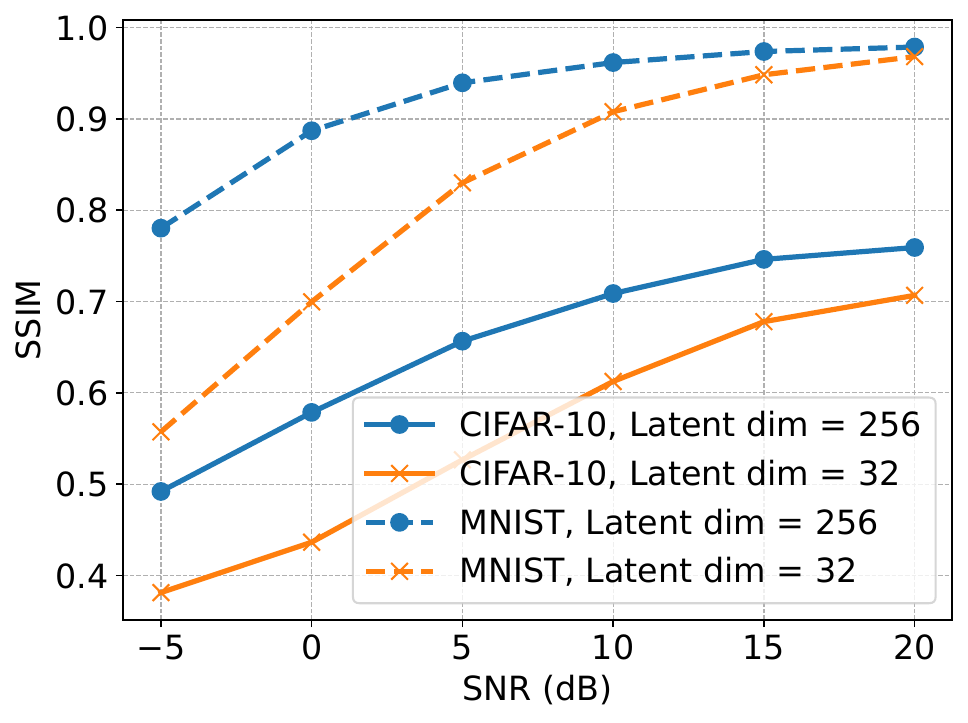}
    \caption{SSIM.}\label{fig:Semd}
  \end{subfigure}

  \caption{Semantic communications performance.}
  \label{fig:Sem}
\end{figure}

Performance of semantic communications is shown in Fig.~\ref{fig:Sem} as a function of SNR for the Alice-Bob channel. Overall, performance improves with SNR. MNIST has lower visual complexity and stronger class separation, and therefore achieves higher semantic accuracy and reconstruction quality than CIFAR-10 under comparable channel and bandwidth conditions. Next, the latent dimension is varied from 32 to 256 to emulate different bandwidth budgets. A higher latent dimension corresponds to more channel uses per image, that is, reduced compression, and improves semantic accuracy, PSNR, and SSIM for both MNIST and CIFAR-10 by supporting more robust discriminative and reconstructive feature encoding under fading and noise.

\section{Privacy-Preserving Semantic Communications Against Eavesdropper \label{sec:privacy}}
\subsection{Threat Model and Eavesdropper Architecture}
We introduce an \emph{unintended receiver} (Eve) that observes the transmitted latent representation through its own Rayleigh fading plus AWGN channel, potentially at a different SNR than Bob, as shown in Fig.~\ref{fig:SysMod}. Eve, equipped with a single antenna, aims to train a strong decoder from the received representations under its channel conditions. Since Eve does not control the transmitter, it has no encoder; instead, it trains a \emph{semantic classifier} that maps the received representation to task labels. We assume Eve uses the same semantic-branch architecture as Bob and, like Bob, performs decoding without explicit CSI as an input. Eve's objective is to exploit whatever task-relevant information remains in the transmitted representation after corruption by its channel.

\subsection{Min-max Training Formulation for Privacy Protection}
Privacy is imposed by introducing \emph{competing objectives} between the legitimate link and the eavesdropper. Eve is trained to strengthen semantic inference by minimizing its loss on intercepted representations. Alice's encoder and Bob's decoders are then optimized with a \emph{dual goal}: maintain Bob's semantic accuracy and reconstruction quality while reducing Eve's ability to infer task-relevant information. Bob is trained with a weighted sum of multi-task losses, using CCE for the semantic task and MSE with an SSIM-based term for reconstruction quality. Privacy pressure is introduced by adding an additional loss term (weighted by a privacy protection weight $w_P$) that penalizes Eve's semantic success, implemented by maximizing Eve's CCE for semantic leakage (equivalently, subtracting Eve's classification loss from the legitimate objective). The privacy protection weight $w_P$ governs the privacy-utility tradeoff, with larger values enforcing stronger semantic privacy at the cost of potential utility loss under tight bandwidth or adverse channel conditions.

The \emph{min-max} interaction is realized via alternating updates. In the first phase, Eve updates its decoder to better infer semantics from the current transmitted representation while Alice's encoder is held fixed. In the second phase, Alice's encoder and Bob's decoders are updated to improve Bob's semantic and reconstruction objectives while reducing Eve's inference capability, with Eve's parameters held fixed and the privacy term computed from Eve's CCE for semantic leakage based on its received signal.  

Repeating these two phases over many epochs yields an adaptive \emph{min-max game} in which Eve continually strengthens its decoder and the legitimate system continually counters. Alternating updates mitigate early dominance by either side and give a direct operational interpretation: privacy is enforced by optimizing the legitimate representation against a continually improving eavesdropper across a range of Eve SNR conditions.

During design and training, Alice and Bob are trained against an Eve surrogate via alternating updates to obtain a conservative privacy assessment, with channel SNRs randomized over a broad range to promote robustness across operating conditions.  In a \emph{distributed} operation, training proceeds via forward transmissions from Alice and limited reverse-link feedback from Bob: Alice transmits an encoded latent waveform, Bob decodes a faded and noisy observation, forms multi-task training signals from its supervision, and returns compact feedback (e.g., quantized or periodic latent updates) so Alice can update the encoder without sharing raw data. Privacy is enforced only during training through the Eve surrogate, which observes its own faded and noisy signal under randomized Eve SNR conditions and contributes its inference loss to the adversarial privacy term within the trusted loop. After training, the Alice-Bob link operates standalone without access to Eve, while Eve performs inference independently on passively intercepted signals and semantic leakage is assessed by Eve's task success. 

\subsection{Performance of Privacy Protection with Min-max Training}
\begin{figure}[t]
  \centering

  \begin{subfigure}[t]{0.49\columnwidth}
    \centering
    \includegraphics[width=\linewidth]{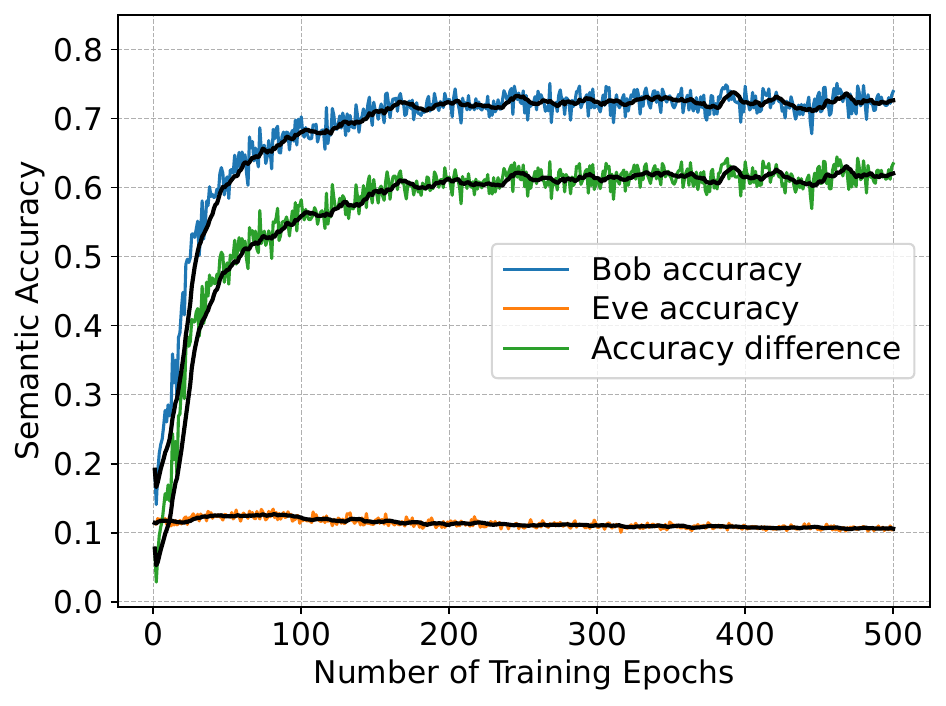}
    \caption{Eve SNR = -5dB.}\label{fig:Evolvemin5}
  \end{subfigure}\hfill
  \begin{subfigure}[t]{0.49\columnwidth}
    \centering
    \includegraphics[width=\linewidth]{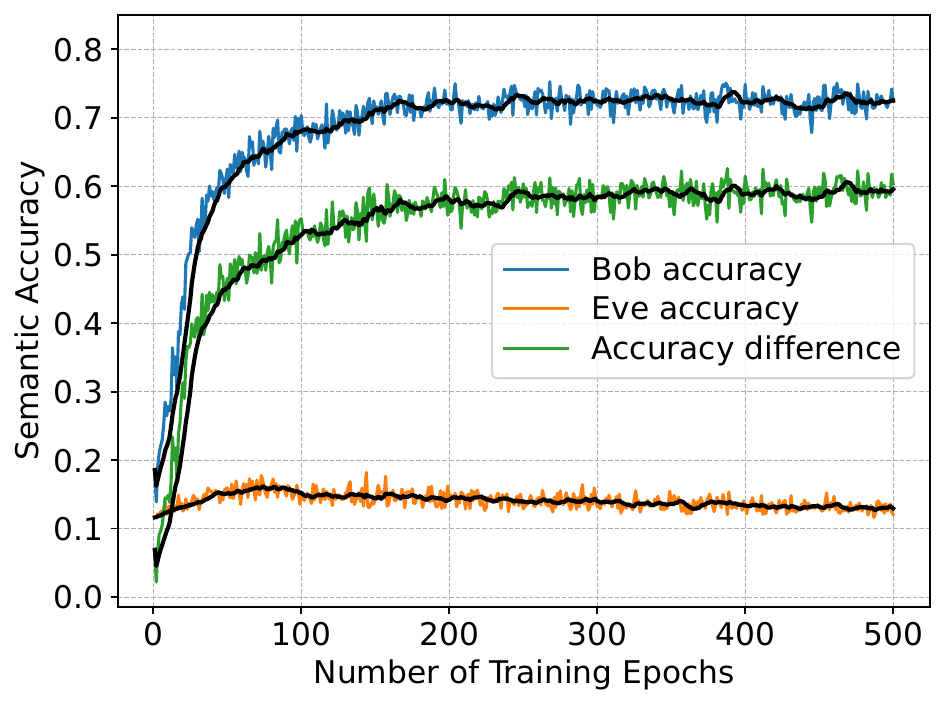}
    \caption{Eve SNR = 0dB.}\label{fig:Evolve0}
  \end{subfigure}

  \vspace{0.6em}

  \begin{subfigure}[t]{0.49\columnwidth}
    \centering
    \includegraphics[width=\linewidth]{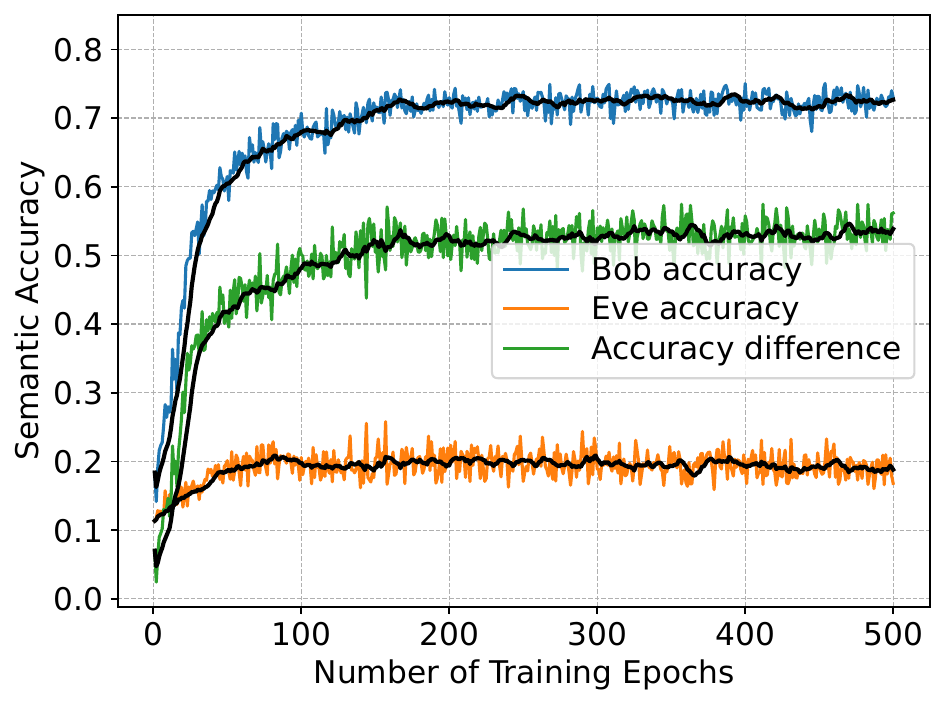}
    \caption{Eve SNR = 5dB.}\label{fig:Evolve5}
  \end{subfigure}\hfill
  \begin{subfigure}[t]{0.49\columnwidth}
    \centering
    \includegraphics[width=\linewidth]{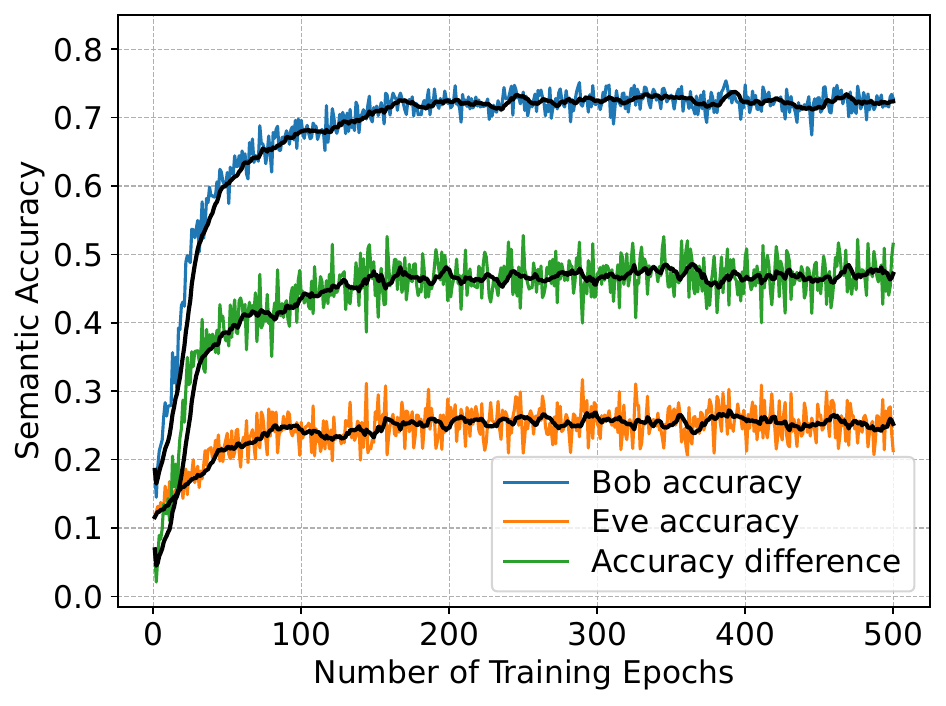}
    \caption{Eve SNR = 10dB.}\label{fig:Evolve10}
  \end{subfigure}

    \caption{{Semantic accuracy during training for different Eve evaluation SNRs (when CIFAR-10 is used and privacy protection weight $w_P$ is 10). }}
  \label{fig:Evolve}
\end{figure}

Fig.~\ref{fig:Evolve} shows the evolution of semantic accuracy for Bob and Eve over epochs when CIFAR-10 is used, the Alice-Bob SNR is 10~dB, and $w_P=10$. Under alternating min-max training, the legitimate Alice-Bob link and Eve both improve from their initial untrained states. As training progresses, Bob's semantic accuracy increases rapidly as the encoder-decoder pair learns a representation that remains reliably decodable over Bob's channel, whereas Eve's semantic accuracy improves more slowly and saturates at relatively low values under privacy pressure. Consequently, the Bob-Eve semantic gap widens before settling into a steady regime. After convergence, accuracies remain stable, and the Bob-Eve gap is largest at lower Eve evaluation SNRs, where channel impairments further limit exploitable information in the intercepted representation.

\begin{figure}[t!]
  \centering

  \begin{subfigure}[t]{0.49\columnwidth}
    \centering
    \includegraphics[width=\linewidth]{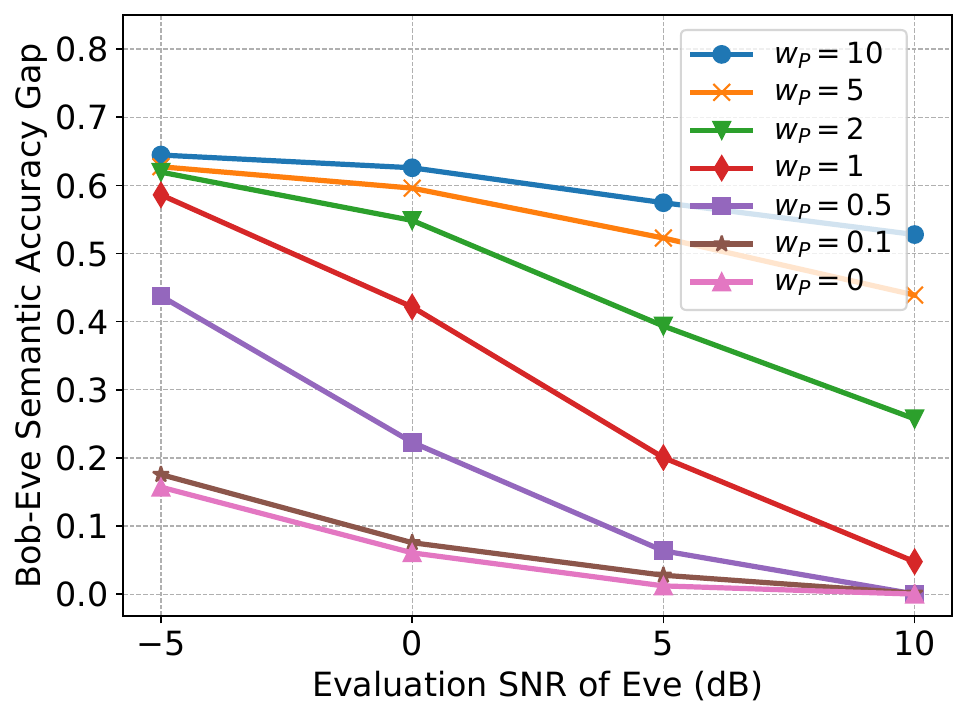}
    \caption{CIFAR-10, latent dimension = 256.}\label{fig:Privacya}
  \end{subfigure}\hfill
  \begin{subfigure}[t]{0.49\columnwidth}
    \centering
    \includegraphics[width=\linewidth]{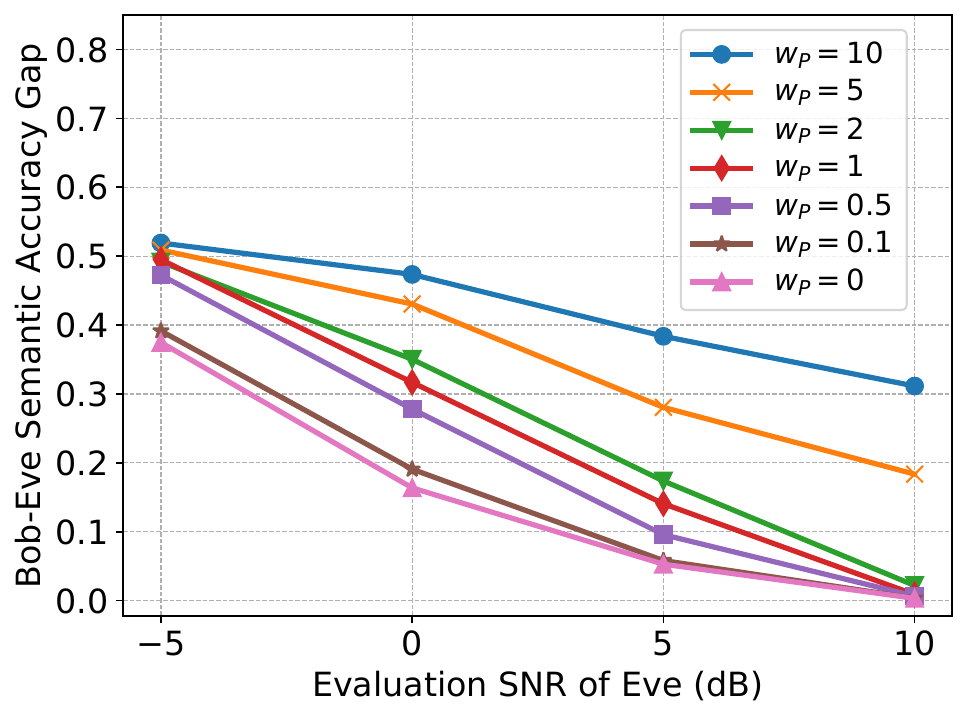}
    \caption{CIFAR-10, latent dimension = 32.}\label{fig:Privacyb}
  \end{subfigure}

  \vspace{0.6em}

  \begin{subfigure}[t]{0.49\columnwidth}
    \centering
    \includegraphics[width=\linewidth]{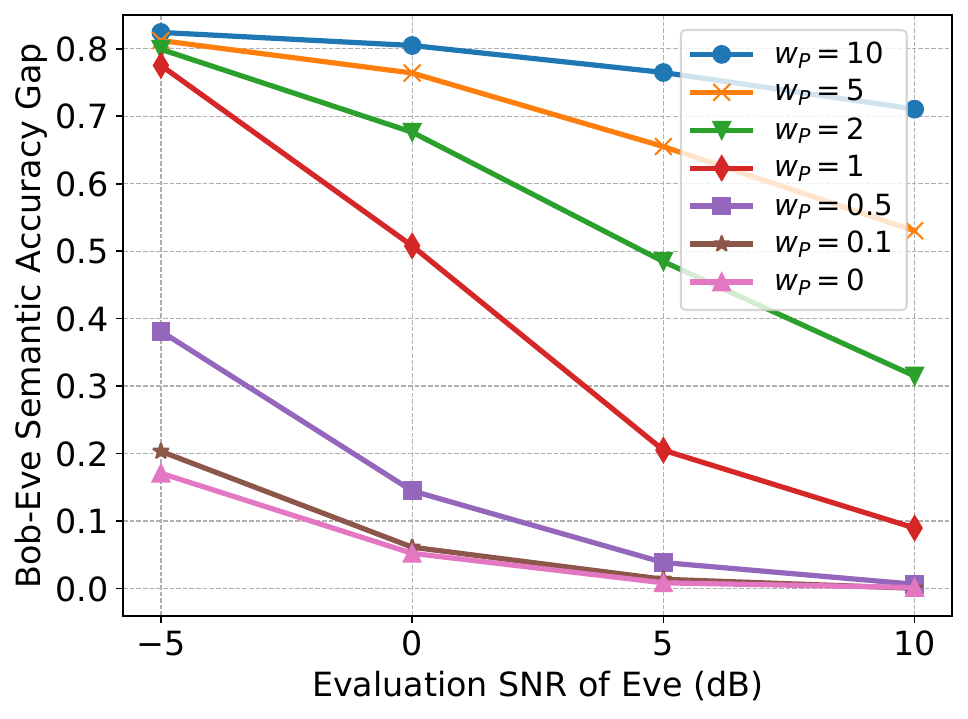}
    \caption{MNIST, latent dimension = 256.}\label{fig:Privacyc}
  \end{subfigure}\hfill
  \begin{subfigure}[t]{0.49\columnwidth}
    \centering
    \includegraphics[width=\linewidth]{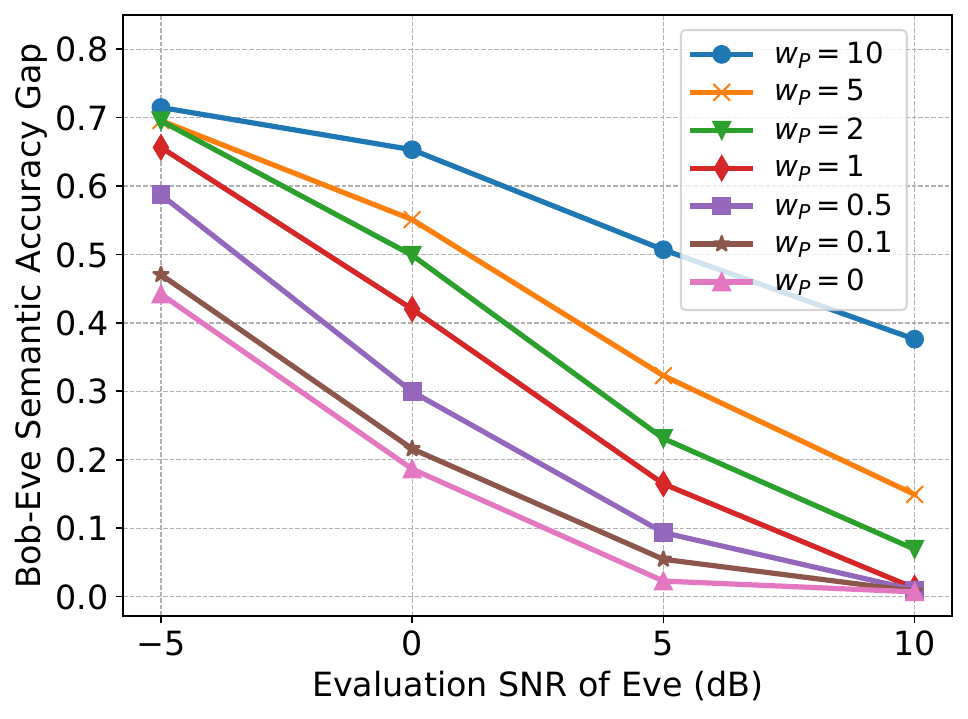}
    \caption{MNIST, latent dimension = 32.}\label{fig:Privacyd}
  \end{subfigure}

  \caption{Protection against eavesdropping for semantic inference.}
  \label{fig:Privacy}
\end{figure}

\emph{With Eve present, the semantic accuracy gap between Bob and Eve serves as an empirical indicator of semantic leakage} (a smaller gap implies greater leakage). Fig.~\ref{fig:Privacy} quantifies the privacy-utility tradeoff by evaluating the Bob-Eve semantic accuracy gap as a function of Eve's evaluation SNR across multiple privacy protection weights $w_P$. Overall, the semantic accuracy gap widens as the Eve SNR decreases for both cases when CIFAR-10 and MNIST images are used. 

Without protection (i.e., when $w_P = 0$), Eve's semantic accuracy closely tracks Bob's, and the semantic accuracy gap rapidly collapses to zero as Eve SNR increases, indicating essentially complete semantic leakage. Larger $w_P$ shifts the semantic accuracy gap upward and helps maintain separation over a broader Eve SNR range. Lower semantic leakage is achieved when larger $w_P$ and larger latent dimension are used. In that case, semantic accuracy is higher for MNIST than CIFAR-10, since MNIST semantics can be preserved for Bob with more compact representations, whereas CIFAR-10 typically requires richer features that are harder to make informative for Bob yet uninformative for Eve.  

Meanwhile, the effect of privacy protection on Bob's reconstruction performance remains limited. For instance, when CIFAR-10 images are used with latent dimension 256, privacy protection changes Bob's PSNR at most by 0.31~dB for $w_P=1$ and 1.55~dB for $w_P=10$ even when Eve evaluation SNR is 10dB, and the corresponding change in SSIM is limited to 0.0042 for $w_P=1$ and 0.0769 for $w_P=10$. These results indicate only a mild fidelity penalty while enforcing semantic privacy against a strong Eve.

\section{Adversarial Perturbations as an Auxiliary Mechanism for Privacy \label{sec:perturbation}}

\subsection{Perturbation Threat Model and Operational Setup}
Privacy can be further strengthened by injecting \emph{adversarial perturbations} that selectively degrade Eve's semantic leakage while trying to preserve Bob's utility. This is realized by a \emph{cooperative jammer} that superimposes a crafted, adversarial perturbation on Alice's signal, as shown in Fig.~\ref{fig:SysMod}. Unlike brute-force jamming, the perturbation is designed to target Eve's learned inference pipeline and reduce Eve's semantic classification accuracy with minimal impact on Bob's semantic and reconstruction tasks.

A key distinction in this approach is that Alice and Bob are trained in a strictly utility-driven manner, \emph{without incorporating Eve into the encoder-decoder objective}. The encoder and Bob's decoders are optimized only for Bob's semantic performance and reconstruction quality over the legitimate channel, with no adversarial term tied to Eve used in the Alice-Bob parameter updates, preserving a clean baseline for end-to-end learning.

Instead, privacy is induced externally via the \emph{cooperative perturbation} mechanism, which provides the only coupling to Eve. Operationally, Alice produces a waveform in latent space, and the cooperative jammer generates a perturbation waveform to disrupt Eve. Eve is modeled using the same learning-based semantic inference architecture as in Sec.~\ref{sec:sem}, operating on its own received representation under its channel conditions.

In \emph{distributed} operation, the cooperative jammer computes and updates the perturbation locally using an Eve model or surrogate together with its own channel state or SNR estimates, aiming to increase Eve's inference error while maintaining Bob's task performance. Because the cooperative jammer's channels to Bob and Eve can differ, the same perturbation can be mild at Bob yet disruptive at Eve, increasing the Bob-Eve semantic gap without changing the core Alice-Bob training objective.

\subsection{Adversarial Perturbation Mechanisms}
In this setup, adversarial perturbations originally developed for attacks are repurposed as a defensive mechanism to degrade Eve's semantic inference from the transmitted representation. The objective is not to disrupt the intended task performance \cite{sagduyu2023task}, but to craft a small cooperative perturbation that selectively degrades Eve's ability to infer task-relevant semantics from intercepted observations. We employ \emph{an untargeted attack} formulation, meaning the perturbation is designed to make any Eve mistake more likely rather than forcing a specific incorrect decision. The eavesdropper's decoder is treated as a differentiable model that maps its received signal to task labels. The perturbation is computed by evaluating how changes in the transmitted waveform affect Eve's inference loss, then choosing a bounded perturbation that increases this loss as much as possible while keeping Bob's semantic and reconstruction performance largely unchanged.

We consider \emph{single-step FGSM and iterative PGD} to generate adversarial perturbations, and vary the number of PGD steps to quantify how perturbation strength impacts semantic leakage. FGSM applies a single fixed size update per transmitted block in the signal direction that most increases Eve's inference loss, efficiently using the full perturbation budget; it is computationally lightweight and latency-friendly, but limited by its one-shot local sensitivity estimate. PGD follows the same principle by starting from the unperturbed signal (or a feasible start), iteratively updating the perturbation to increase Eve's loss, and projecting onto the constraint set after each step. By better tracking nonlinear sensitivities, PGD typically yields stronger disruption than FGSM under the same budget, and more iterations can further improve effectiveness.

Bob and Eve each observe a superposition of the transmissions from Alice and the cooperative jammer, scaled by their respective channels and corrupted by noise. Because the cooperative jammer's channels to Bob and Eve typically differ, the same perturbation can be relatively benign at Bob while remaining disruptive at Eve, providing a practical degree of freedom to increase the Bob-Eve semantic accuracy gap without changing Bob's decoder or the encoder's task-oriented representation learning. The cooperative jammer may use coarse link quality estimates (e.g., SNR) for power control to limit impact at Bob while degrading Eve.

Training keeps Eve adaptive while separating utility learning from protection. Eve updates its classifier with the encoder held fixed under the current signaling and perturbation conditions, while Alice and Bob update only Bob's semantic and reconstruction objectives. Thus, any drop in Eve's performance is attributable to the crafted perturbation rather than changes in the legitimate learning objective, yielding a clear interpretation: Alice and Bob optimize communication quality and the perturbation layer limits Eve's success.

\begin{figure}[t]
  \centering

  \begin{subfigure}[t]{0.49\columnwidth}
    \centering
    \includegraphics[width=\linewidth]{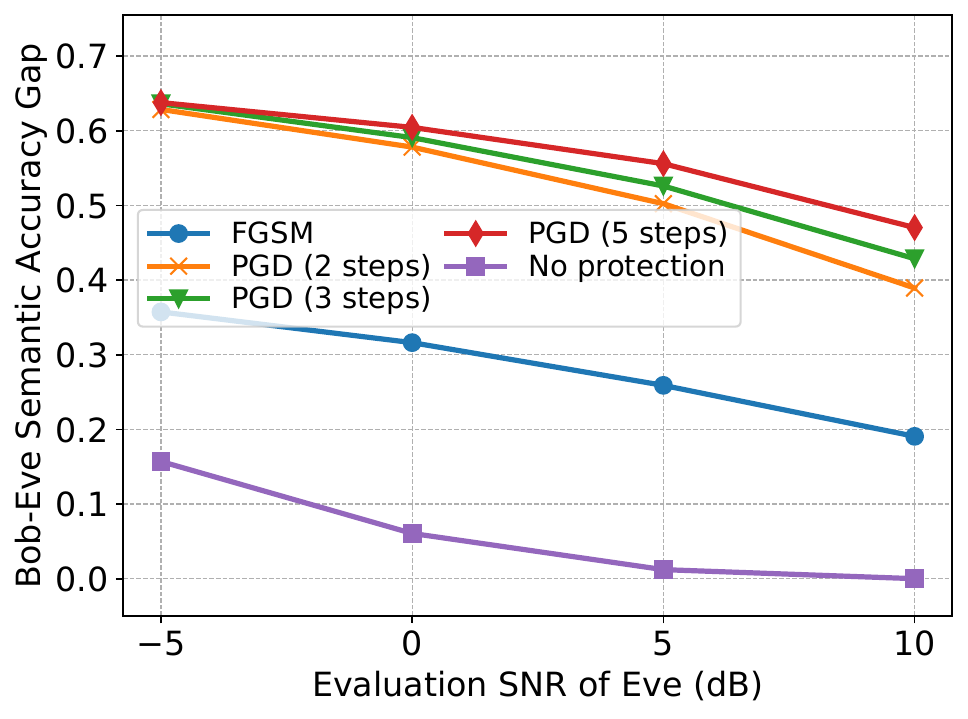}
    \caption{CIFAR-10, latent dimension = 256.}\label{fig:Perturba}
  \end{subfigure}\hfill
  \begin{subfigure}[t]{0.49\columnwidth}
    \centering
    \includegraphics[width=\linewidth]{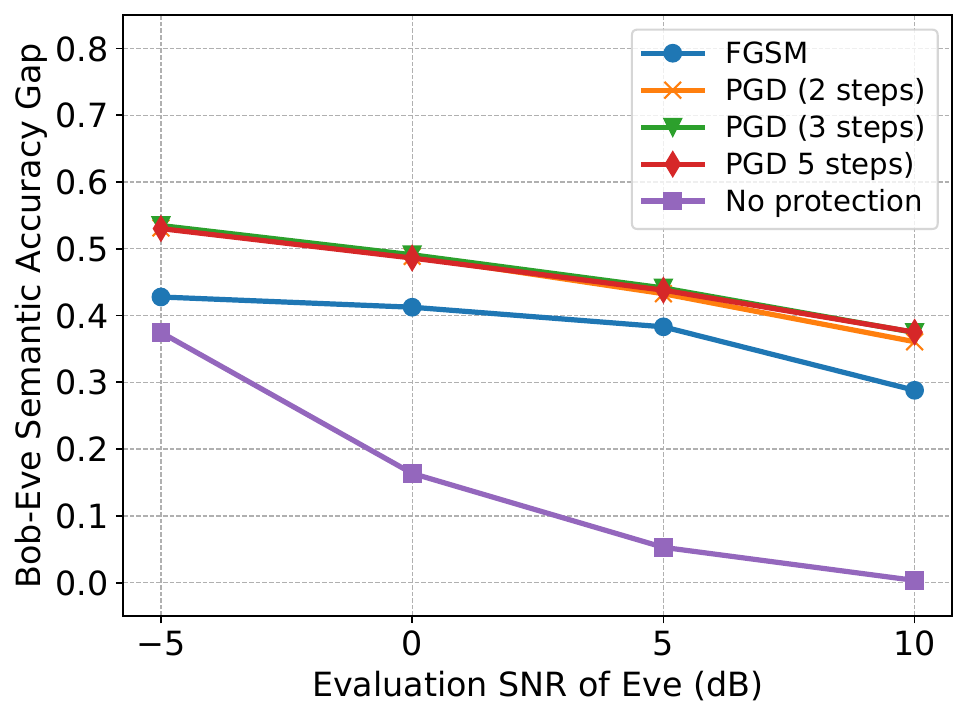}
    \caption{CIFAR-10, latent dimension = 32.}\label{fig:Perturbb}
  \end{subfigure}

  \vspace{0.6em}

  \begin{subfigure}[t]{0.49\columnwidth}
    \centering
    \includegraphics[width=\linewidth]{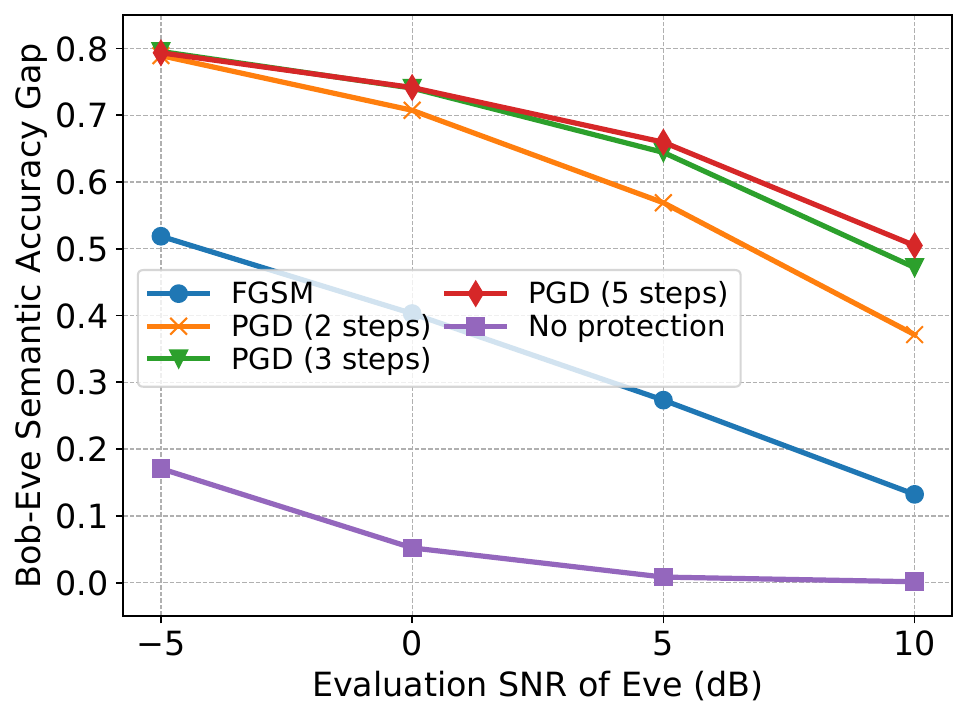}
    \caption{MNIST, latent dimension = 256.}\label{fig:Perturbc}
  \end{subfigure}\hfill
  \begin{subfigure}[t]{0.49\columnwidth}
    \centering
    \includegraphics[width=\linewidth]{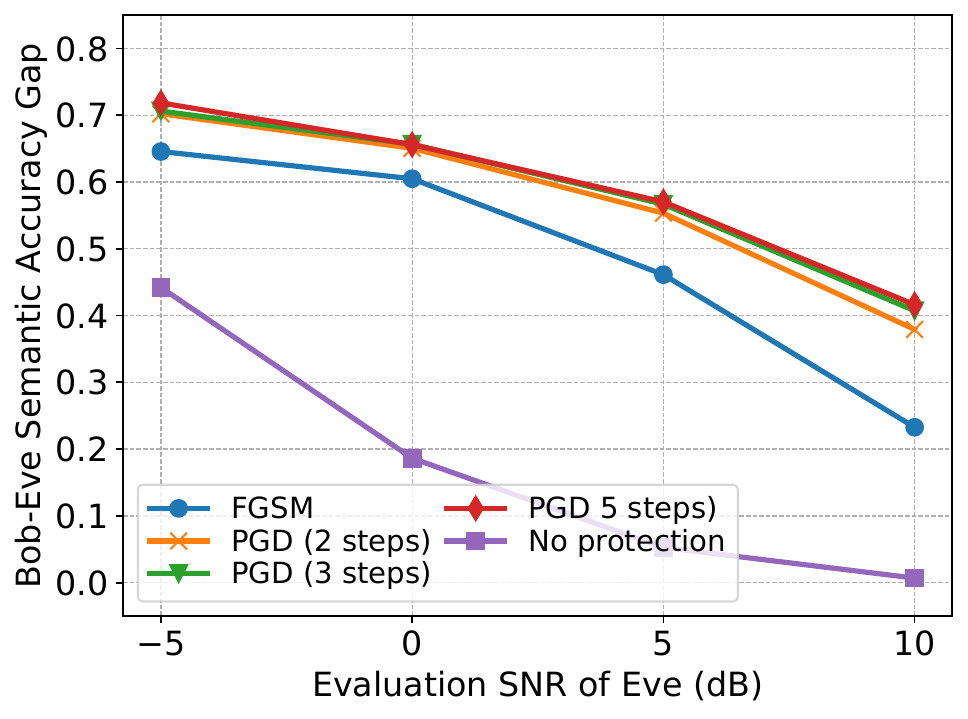}
    \caption{MNIST, latent dimension = 32.}\label{fig:Perturbd}
  \end{subfigure}

  \caption{Privacy protection via adversarial perturbations.}
  \label{fig:Perturb}
\end{figure}

\subsection{Performance of Privacy Protection with Adversarial Perturbations}
Fig.~\ref{fig:Perturb} reports the Bob-Eve semantic accuracy gap versus Eve evaluation SNR under the no-protection baseline and under perturbation-based protection using FGSM and PGD with increasing PGD steps. Across the evaluated Eve-SNR range, both FGSM and PGD increase the semantic accuracy gap significantly relative to the no-protection baseline, indicating reduced semantic leakage under the same core semantic encoder-decoder. 

PGD consistently yields a larger gap than FGSM. Increasing the number of PGD steps further reduces semantic leakage (even when the latent dimension is larger), reflecting the benefit of stronger multi-step optimization of the perturbation under the same utility-preservation constraint at Bob. With a smaller latent dimension, the semantic bottleneck leaves limited flexibility in the transmitted representation, so the cooperative jammer has fewer effective directions to further confuse Eve without also affecting Bob, and additional PGD iterations quickly saturate.  The semantic accuracy gap decreases as Eve SNR increases, but multi-step PGD preserves a larger separation at moderate-to-high Eve SNRs than single-step FGSM, offering an additional mechanism for strengthening semantic privacy when the eavesdropper operates under favorable channel conditions. At the same time, the impact on Bob's semantic inference and reconstruction performance remains limited, as also observed in Sec.~\ref{sec:privacy}.

\section{Future Directions \label{sec:future}}
The results suggest that privacy can be treated as an explicit, controllable objective in learned semantic communications without sacrificing the legitimate link's core task performance across a range of channel conditions and bandwidth budgets. Several directions can broaden applicability, strengthen robustness, and improve interpretability by extending the framework to richer tasks and leakage notions, stronger and more adaptive adversaries, and practical system constraints that govern deployment and real-time operation.

\begin{itemize}
\item \emph{Richer downstream tasks and leakage notions:}
Move beyond single-label classification to segmentation, retrieval, and multi-label recognition, and assess leakage beyond top-1 accuracy (e.g., attribute inference, membership inference, feature inversion, and task-conditioned mutual information surrogates).

\item \emph{Stronger eavesdroppers:}
Consider adversaries with side information, ensembles, self-supervised pretraining, and transfer learning, and evaluate robustness under distribution shift, non-stationary channels, and mismatched hardware impairments, including adaptive decoders with online updates.

\item \emph{Eavesdropper reconstruction as an additional leakage channel:}
Equip Eve with an additional decoder for reconstruction in addition to semantic inference and study how joint objectives change the privacy-utility tradeoff; likewise, design perturbations that target Eve reconstruction alone or jointly with inference while preserving Bob performance.

\item \emph{Perturbation design and robustness:}
Study alternative methods (e.g., DeepFool and Carlini-Wagner), budgets, step sizes, and iteration counts under practical constraints (power, spectral mask, latency), and evaluate robustness to countermeasures such as adversarial training, randomized smoothing, and denoising.

\item \emph{Multi-user and multi-antenna generalizations:}
Extend to MIMO and multi-user settings that exploit spatial degrees of freedom and interference structure, including heterogeneous privacy requirements, user scheduling, and fairness under shared spectrum access.

\item \emph{Training stability and efficiency:}
Develop competitive optimization schedules (e.g., curriculum over Eve SNR, gradient penalty regularization, early stopping) to reduce collapse modes and lower sweep cost over latent dimension, SNR, and protection weights; use lightweight Eve surrogates for rapid design space exploration.

\item \emph{Connecting empirical protection to formal metrics:}
Relate learned adversary evaluations to information theoretic or statistical leakage measures (e.g., hypothesis testing bounds and rate distortion privacy tradeoffs) to enable clearer guarantees and more comparable benchmarks.

\item \emph{Covertness and traffic analysis resilience:}
Incorporate covertness objectives to reduce detectability and traffic analysis leakage (e.g., statistical indistinguishability from noise and other signals) and study joint privacy-covertness tradeoffs.

\item \emph{Higher-layer network extensions and cross-layer design}: Extend the physical-layer semantic privacy mechanism to cross-layer designs that jointly optimize semantic compression with scheduling, retransmission, routing, congestion control, and resource allocation under latency-reliability constraints, while quantifying semantic leakage under traffic analysis and metadata exposure.

\item \emph{System-level integration and deployment:}
Develop and demonstrate end-to-end radio implementation and fielded deployment under practical constraints (e.g., synchronization, timing/frequency offset and phase-noise, quantization, and real-time compute and power limits), and derive guidelines for selecting latent dimension and protection weights to meet mission-driven utility targets within explicit risk budgets.
\end{itemize}

\section{Conclusion \label{sec:conclusion}}
This paper presented a unified framework for privacy-preserving semantic communications that combines multi-task learning, competitive min-max optimization, and perturbation-based protection against an adaptive eavesdropper. By explicitly modeling semantic leakage through a learning-based eavesdropper, privacy becomes a controllable design objective, not an implicit byproduct of waveform design. The legitimate transmitter-receiver pair learns an end-to-end representation supporting semantic inference and reconstruction, while privacy is enforced either through competitive training or via an external cooperative perturbation layer that preserves the core utility objective. Results show that training against an adaptive eavesdropper preserves high semantic accuracy at the legitimate receiver while reducing the eavesdropper's inference success, enabling a tunable privacy-utility tradeoff. The perturbation mechanism further reduces leakage even when the legitimate link is trained purely for utility, providing a complementary, deployment-friendly protection layer. The reduction in semantic leakage is characterized across operating regimes as a function of channel conditions, the latent dimension (as a proxy for channel uses and compression), the privacy protection weight, and the adversarial perturbation method. Looking ahead, these complementary, tunable mechanisms point toward semantic communication systems that can adapt protection online to evolving tasks, channel dynamics, and adversary capabilities, and deliver task utility with quantifiable, context-aware privacy protection in real-world deployments.

\bibliographystyle{IEEEtran}
\bibliography{references}

\end{document}